\documentclass[12pt,english]{elsarticle}
\usepackage[T1]{fontenc}
\usepackage[latin9]{inputenc}
\usepackage{amstext}
\usepackage{amssymb}
\usepackage{graphicx}
\usepackage{esint}

\makeatletter
\journal{ArXiv}

\makeatother

\usepackage{babel}
\begin{document}

\title{An OPERA inspired classical model reproducing superluminal velocities}

\author{Bogus\l{}aw Broda}

\ead{bobroda@uni.lodz.pl}

\ead{http://merlin.fic.uni.lodz.pl/kft/people/BBroda}

\address{Department of Theoretical Physics, University of \L{}\'od\'{z}, Pomorska
149/153, PL--90-236~\L{}\'od\'{z}, Poland}
\begin{abstract}
In the context of the sensational results concerning superluminal
velocities, announced recently by the OPERA Collaboration, we have
proposed a classical model yielding a statistically calculated measured
velocity of a beam, higher than the velocity of the particles constituting
the beam. The two key elements of our model, necessary and sufficient
to obtain this curious result, are a time-dependent {}``transmission''
function and statistical method of the maximum-likelihood estimation. \end{abstract}
\maketitle
\def\citet{\cite}

Inspired by the amazing results of the OPERA Collaboration \citet{Opera},
claiming that the velocity of light $c$ has been beaten by a beam
of neutrinos, we propose a simple classical model, which yields the
curious effect of a seeming increase of {}``effective'' velocity
(see \citet{Gilles}, for an earlier independent approach, and also
\citet{Alicki}, for a wave version). There are dozens of papers,
which have recently appeared, adopting various attitudes towards the
results presented by the OPERA Collaboration, for example \citet{CibRem,BYYY,PfeWoh,Saridakis,ABP}
(see also an example of an older work on superluminal velocities \citet{Alfaro}).
The attitude of our paper is definitely skeptical. In view of the
fact that we have found a ({}``mathematical'') model providing an
{}``artificial'' increase of real velocity, we are forced to put
the results announced by the OPERA Collaboration in doubt. Moreover,
we have also proposed a simple working example, such that appropriately
fitting its parameters one can simulate the controversial results
of the OPERA Collaboration. Strictly speaking, our paper does not
indisputably invalidates the conclusions drawn by the OPERA Collaboration,
but it seriously weakens their argumentation, indicating a logical
gap in their reasoning.

Our model is purely classical and dynamics free. No new physics, nor
quantum mechanics, nor even (classical) wave mechanics is involved.
Only standard classical kinematics notions as well as the statistical
method of the maximum-likelihood estimation (MLE) are used in our
approach.

The assumptions of our model are the following. A spatially homogeneous,
lasting the period $T$, beam of classical particles ({}``extraction'')
moving with a constant speed $u$ ($\leqq c$) travels from a source
to a detector. The distance between the source and the detector is
$d$. The probability density function (PDF) of the time of emission
of the particles within the duration $T$ of production of the beam
is given by the function $w(t)$. In an ideal situation (none of the
particles is lost) we would obtain, according to classical kinematics,
the measured data waveform $y(t)=w(t-t_{0})$, where $t_{0}=d/u$.
Now, let as suppose that the fraction of the particles emitted, measured
by the detector, due to some physical mechanism, is given by the (non-negative)
{}``transmission'' function
\begin{equation}
f\left(t\right)\equiv\frac{N_{d}\left(t\right)}{N_{e}(t)},\label{eq:fraction}
\end{equation}
where $N_{e}\left(t\right)$ is the number of the particles emitted
at the time instant $t,$ and $N_{d}\left(t\right)$ is the number
of the particles detected, which have been sent at the same time instant
$t$. In other words, in general, not all particles emitted are detected
($f(t)<1$) --- obvious, and moreover $f(t)\neq\textrm{const}$ ---
conceivable. For simplicity, we will assume that the numbers of the
particles, $N_{d}(t)$ and $N_{e}(t)$, are so large that we are allowed
to use a continuous approximation. Then, the transmission function
$f(t)$ is a (continuous) function satisfying the condition $0\leqq f(t)\leqq1$.

To be able to draw conclusions from experimental data, we should implement
some statistical methodology. To be so precise as possible, in our
approach, we adopt the method of the MLE, as the OPERA Collaboration
has done \citet{Opera}. In the framework of the MLE, we introduce
the likelihood function $L$. The logarithm $l$ of $L$ is given
by the formula 
\begin{equation}
l(\delta t)\equiv\log\left[L(\delta t)\right]\equiv\sum_{j}\log\left[w(t_{j}+\delta t)\right],\label{eq:genlikelihood}
\end{equation}
where $t_{j}$ are the time instants corresponding to the measurement
events at the detector, and the time deviation $\delta t$ we are
interested in (see \citet{Opera}, for details)  provides the maximum
of $l$ (and also of $L$). As the numbers of the measured events
$t_{j}$ are large, in our continuous approach, instead of summation
in (\ref{eq:genlikelihood}), we should use integration with an appropriate
integration measure $z(t)dt$, where $z(t)$ represents the time distribution
of the experimental events detected by the OPERA. In fact, $z(t)$
is determined by product of two factors. The first factor, $y$, is
proportional to the number of the particles sent, i.e. $y(t+t_{0})=w(t)$,
and the second one is proportional to the transmission function $f(t)$.
Then,
\begin{equation}
z(t)=f(t)y(t+t_{0})=f(t)w(t).\label{eq:defz}
\end{equation}

Finally, 
\begin{equation}
l(\delta t)=\int\log\left[w(t+\delta t)\right]f(t)w(t)dt.\label{eq:conlikelihood}
\end{equation}

\begin{figure}
\noindent \begin{centering}
\includegraphics[scale=0.8]{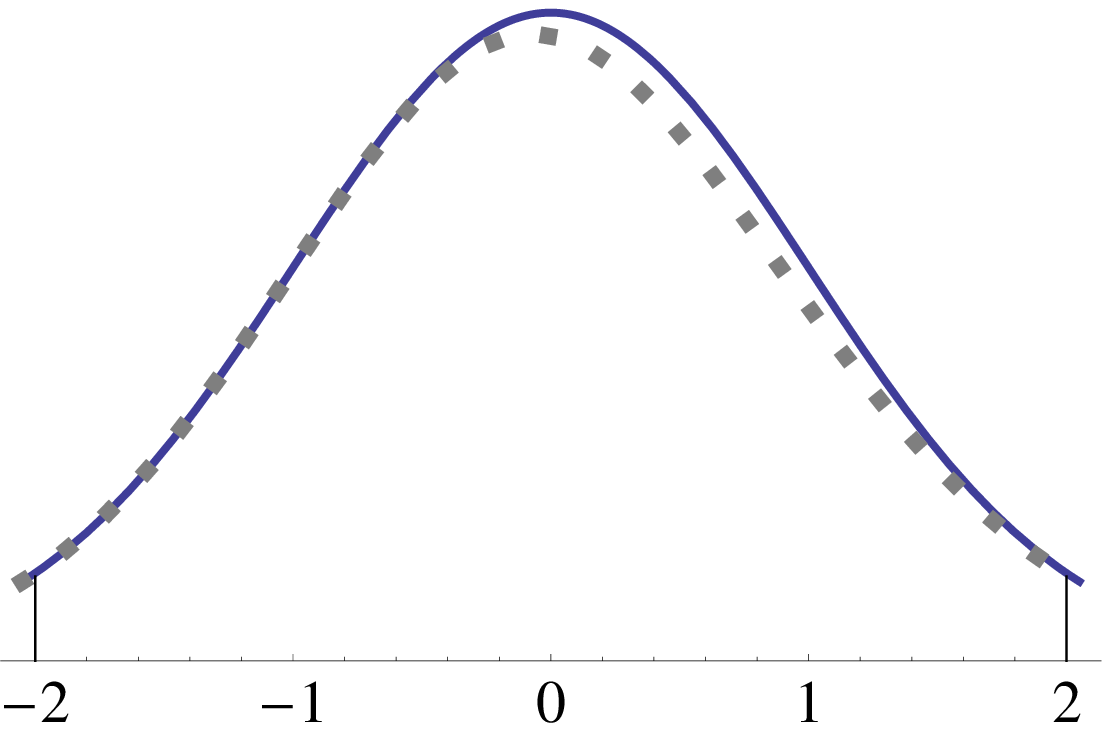}
\par\end{centering}

\caption{\label{fig:2curves}}

\noindent \centering{}{\small The presented specific example is intended
to {}``mathematically'' qualitatively simulate the OPERA experiment,
yielding the time deviation $\delta t\approx+75.5\,\textrm{ns}$.
The solid curve represents the probability density function (PDF)
$w(t)$, whereas the dashed one corresponds to the time distribution
of the {}``detected experimental data'' $z(t)$.} 
\end{figure}

To demonstrate that our idea actually works, we propose a specific
example. The parameters of the example are so fitted that it yields
the time deviation $\delta t\approx+75.5\,\textrm{ns}$ (for comparison,
the OPERA result is $60.7\,\textrm{ns}$). For calculational simplicity,
we assume the Gaussian form of the PDF (see the solid line in Fig.\ref{fig:2curves}),
\begin{equation}
w(t)=\exp\left(-\frac{t^{2}}{2}\right),\label{eq:gauss}
\end{equation}
as well as the Gaussian form of of the transmission function, 
\begin{equation}
f(t)=1-\frac{1}{10}\exp\left[-\left(t-1\right)^{2}\right],\label{eq:nontrivialtransmission}
\end{equation}
where the one tenth in front of the exponent is reminiscent of {}``$10\%$
variation'' in \citet{Gilles}. The time distribution of the {}``detected
experimental data'' $z(t)$, corresponding to $w(t)$ of the form
(\ref{eq:gauss}) and to the transmission function $f(t)$ of the
form (\ref{eq:nontrivialtransmission}) is presented in Fig.\ref{fig:2curves}
by the dashed curve. 
In this (doubly) Gaussian case it is even possible to solve the problem
analytically (see Appendix), but for our purposes a numerical value
will do. It appears, that the maximum of (\ref{eq:conlikelihood})
is attained for the time deviation 
\begin{equation}
\delta t\approx+0.0288\,\textrm{s.d}.,\label{eq:numdelt}
\end{equation}
where s.d.\ means the standard deviation. One can easily translate
the dimensionless (\ref{eq:numdelt}) into a dimensionfull entity.
In the OPERA experiment $T\approx10.5\,\mu\textrm{s}$, whereas in
our example we can reasonably assume, for definiteness, that $T$
equals (in dimensionless units, or in the units of the standard deviation)
twice the two standard deviations, i.e.\  $T\approx2\cdot2=4$ (see
two vertical intervals in Fig.\ref{fig:2curves}). Then, the dimensionfull
time deviation corresponding to (\ref{eq:numdelt}) is 
\begin{equation}
\delta t\approx\frac{0.0288\cdot10.5}{4}\,\mu\textrm{s}\approx75.5\,\textrm{ns}\approx60.7\,\textrm{ns}\,(\textrm{OPERA}).\label{eq:unitdelt}
\end{equation}

In particular, our analysis confirms the findings of \citet{Gilles}
that the time deviation $\delta t$ is independent of the distance
$d$, but depends on the shape of the beam. Obviously, the numerical
coincidence (\ref{eq:unitdelt}) has only an illustrative purpose.
The non-zero time deviation $\delta t$ directly implies a modified
velocity $v$ according to the elementary formula
\begin{equation}
\frac{v-u}{u}\approx\frac{\delta t}{t_{0}}.\label{eq:velocities}
\end{equation}
In our model the sign of $\delta t$ (positive in our example) depends
on details of the form of the transmission function $f(t)$. E.g.,
reversing the sign in front of the exponent in (\ref{eq:nontrivialtransmission})
reverses the sign of $\delta t$.

We would like to stress that our specific example is not supposed
to mimic a real situation in the OPERA experiment, but only to demonstrate
that it is, in principle, possible to easily come to its conclusions
without any reference to superluminal particles and/or some exotic
phenomena. Any considerations concerning a possible physical mechanism
governing the time-dependence of the {}``transmission'' function
$f(t)$ are outside the scope of our paper. We can only speculate
that $f(t)\neq\textrm{const}$ could be a property of the source,
or of the detector or it could follow from interactions in the Earth's
crust. Moreover, measurement errors, always encountered in real experiments,
do not enter our considerations as they have nothing to do with the
discussed effect.

It is also possible to intuitively explain the non-zero value of $\delta t$.
Namely, deforming the shape of the PDF $w(t)$ with an appropriate
(time asymmetric) transmission function $f(t)$ shifts a portion (its
{}``upper'' part) of the {}``waveform'' $w(t)$ forward or backward,
yielding positive or negative $\delta t$, respectively, which is
next erroneously interpreted by the method of the MLE as a {}``real''
time deviation. In a sense, our approach is a {}``corpuscular''
analog of the well-known situation of {}``superluminal'' velocity
of light in material media \citet{BLB}, where to some extent, the
role of $f(t)$ plays dispersion.

In conclusion, we would like to emphasize that there are two elements
of our model, responsible for the curious effect of the superluminal
velocity: the time-dependent transmission function $f(t)$, which
should {}``favor'' leading edges and {}``disfavor'' trailing edges
of the beam, and statistical methodology, erroneously assuming the
method of the MLE. This is why it is conceivable that the paradoxical
results of the OPERA Collaboration could be possibly avoided upon
another approach, e.g.\ an approach giving preference to the front
velocity or to another statistical method. Thus, no superluminal particles
are necessary to explain the {}``superluminal'' velocities derived
by the OPERA Collaboration. Moreover, independently of the further
fate of the conclusions drawn by the OPERA Collaboration (acceptance
or refutation), we have demonstrated that standard statistical methods,
e.g.\ MLE, should be used with great care, as they can provide erroneous
output. Therefore, the results of our paper do not rely on the final
solution of the OPERA paradox.

\appendix

\section{Analytical form}

For the Reader's convenience, we present here analytical form of the
expressions used in the main part of the paper. For $w(t)$ given
by (\ref{eq:gauss}) and $f(t)$ given by (\ref{eq:nontrivialtransmission}),
we obtain the integral $\left(\ref{eq:conlikelihood}\right)$

\[
l(\delta t)=\intop_{-\infty}^{+\infty}\left[-\frac{\left(t+\delta t\right)^{2}}{2}\right]\left[1-\frac{e^{-\left(t-1\right)^{2}}}{10}\right]\exp\left(-\frac{t^{2}}{2}\right)dt
\]
\begin{equation}
=\frac{\sqrt{\pi}\left[\sqrt{3}\left(7+12\,\delta t+9\,\delta t^{2}\right)-270\sqrt[3]{e}\left(1+\delta t^{2}\right)\right]}{270\sqrt{2}\sqrt[\,3]{e}}.\label{eq:analytint}
\end{equation}

The maximum of $l(\delta t)$ is determined by its derivative

\begin{equation}
l'\left(\delta t\right)=\frac{\sqrt{\pi}\left[\sqrt{3}\left(12+18\,\delta t\right)-540\sqrt[3]{e}\,\delta t\right]}{270\sqrt{2}\sqrt[\,3]{e}},\label{eq:analytder}
\end{equation}

and is attained for

\begin{equation}
\delta t=\frac{2}{30\sqrt{3}\sqrt[\,3]{e}-3}.\label{eq:analytdt}
\end{equation}

\end{document}